\documentclass[aps,twocolumn,showpacs,preprintnumbers,amssymb,prl]{revtex4}

\newcommand{\be}{\begin{eqnarray}}
\newcommand{\ee}{\end{eqnarray}}

\newcommand{\eins}{\mbox{$1 \hspace{-1.0mm}  {\bf l}$}}
\newcommand{\ket}[1]{\left|{#1}\right\rangle}
\newcommand{\bra}[1]{\left\langle{#1}\right|}

\newcommand{\ketbrad}[1]{\left|{#1}\rangle\!\langle{#1}\right|}
\newcommand{\ketbra}[2]{\left|{#1}\rangle\!\langle{#2}\right|}
\newcommand{\mean}[1]{\langle{#1}\rangle}
\def\bea{\begin{eqnarray}}
\def\eea{\end{eqnarray}}

\def\C{\hbox{$\mit I$\kern-.7em$\mit C$}}
\def\N{\hbox{$\mit I$\kern-.3em$\mit N$}}

\def\tr{{\rm tr}}

\usepackage{dcolumn} 
\usepackage{bm} 

\usepackage{epsfig}

\usepackage{amsmath,amsfonts,amssymb,graphics,graphicx,epsfig,color,times,bbm}

\begin{document}

\title{Entanglement and decoherence in spin gases}

\author{ J. Calsamiglia$^{1}$, L. Hartmann$^{1}$, W. D{\"u}r$^{1,2}$, and H.-J. Briegel$^{1,2}$}

\affiliation{$^1$ Institut f{\"u}r Theoretische Physik,
Universit{\"a}t Innsbruck,
Technikerstra{\ss}e 25, A-6020 Innsbruck, Austria\\
$^2$ Institut f\"ur Quantenoptik und Quanteninformation der
\"Osterreichischen Akademie der Wissenschaften, Innsbruck, Austria.}
\date{\today}

\begin{abstract}

We study the dynamics of entanglement in \emph{spin gases}. A spin gas consists of a (large) number of interacting particles whose random motion is described classically while their internal degrees of freedom  are described quantum-mechanically.   We determine the entanglement that occurs naturally in such systems for specific types of quantum interactions. At the same time, these systems provide microscopic models for non--Markovian decoherence:
the interaction of  a group of particles  with other particles belonging
to a background gas are treated exactly, and differences between collective and non--collective
decoherence processes are studied.  We give quantitative results for the  Boltzmann gas and also for a lattice gas, which could be realized by neutral atoms hopping in an optical lattice.
These models can be simulated efficiently for  systems of mesoscopic sizes ($N \sim 10^5$). 
\end{abstract}

\pacs{75.10.Pq, 03.67.Mn, 03.65.Ud, 03.67.-a}

\maketitle

We study the entanglement properties of \emph{spin gases}. A spin gas is a system of interacting spins (or  qubits) where the coupling strengths between the spins are stochastic functions of time. A system that could serve as a textbook example of a spin gas is the semi-quantal Boltzmann gas, where each particle carries an internal (two-level) quantum degree of freedom. During a collision of two particles, the internal degrees of freedom interact and can become entangled. The statistics of the collisions, described by kinetic gas theory, leads to randomly fluctuating coupling strengths between the spins. A similar situation arises in systems where the gas particles do not move freely in space, but are confined to lattice--sites, between which they can classically hop with a certain probability.  It is an intriguing question, how the evolution of the quantum state of the system is determined by the underlying classical thermodynamics of the gas. What kind of entanglement is created in the gas and at which rate? How is the equilibrium state of the gas characterized in terms of its entanglement? In this paper we will give an answer to these and to other questions, which make  the study of spin gases 
interesting both from the perspective of thermodynamics and of quantum information.

Spin gases differ from spin lattices in that the coupling strengths have  no translational symmetry and evolve in time. Spin gases are more closely related to spin glasses, which have random, albeit static,  couplings between the spins. Although there has been much recent work investigating e.g. the role of entanglement in quantum critical phenomena \cite{arnesen01,osborne02,osterloh02, vidal03, vidal03b,Ve03},  there is little theoretical work studying disordered quantum systems such as spin glasses and spin gases from a similar perspective (see, however, \cite{damski03}).

We give a simple, and yet realistic, collision model for
the particles carrying the spins,  and study the quantum mechanical states that emerge from the dynamics of such a spin gas, which we also refer to  as a semi-quantal gas.
In full generality, this problem seems intractable for various reasons:
the description of a many-body quantum state usually requires exponentially large resources; strong interactions do not allow for a perturbative treatment; random interactions prevent  the  appearance of symmetries and the corresponding reduction of the effective number of degrees of freedom; finally, the restriction to low--energy eigenspaces, suitable for the study of ground--state or low--temperature properties, can not  be applied  here to the study of dynamics.  
State of the art numerical methods ---such as density matrix renormalization group  \cite{schollwock04}---  are limited to systems of moderate size (up to a few hundred particles) with bounded amount of entanglement  \cite{vidal03}  (e.g. non--critical spin chains) or an amount  of entanglement that scales at most with the surface of the block of spins \cite{verstraete04} (e.g. some 2-dimensional spin lattices). In disordered quantum systems with random interactions, like the spin gas studied here, entanglement will typically increase with the volume of the block, which implies that these numerical methods can not be applied.
Nevertheless, for Ising (or, more generally, commuting two-body) interactions we can compute the full dynamics of the many-body system exactly and efficiently.
Semi--quantal gases are not only toy
models of theoretical interest, but could be experimentally realized
even with present--day technology and existing setups.

In this paper we characterize the states that arise in spin gases, and calculate their 
expected entanglement properties. We point out that we describe the system by a pure state, 
albeit with random coefficients, and calculate expected properties by averaging them over different realizations of the state.  The formalism covers a  range of different models of spin gases and allows us to identify many of their quantum features.   
For instance, we can describe the quantum dynamics of a particular set of (probe) particles prepared in various initial states.The gas provides a microscopic  model for non-Markovian decoherence \cite{ziman04} that can be treated exactly  even for  large system sizes. 
The paper is structured as follows: we first introduce the formal framework to study spin-gases, describe general entanglement properties, and then treat two specific models of a spin gas. We present  analytical results for semi-quantal Boltzmann gas,  and  numerical results for a lattice gas.

\pagebreak

{\bf Formal preliminaries.} We consider a situation where $N$ particles move along some classical trajectories  ${\bm r}_k(t)$, while their quantum degrees of freedom interact according to a
distance-- and time--dependent Hamiltonian
\be
\label{H}
H(t)=\sum_{k<l} g({\bm r}_k(t),{\bm r}_l(t)) H^{(kl)},
\ee
where $g({\bm r}_k(t),{\bm r}_l(t))$ is some function depending on the particular two--body interaction. 
We restrict ourselves to specific types of interactions, namely those where all $H^{(kl)}$
commute.  Consequently, we find that after a time $t$ the initial state  $|\Psi_0\rangle$ evolves to
\be
|\Psi_t\rangle =U_t \ket{\Psi_0}= \prod_{k>l}
{U}^{(kl)}(\varphi_{kl}(t)) |\Psi_0\rangle,
\label{eq:Us}
\ee 
with $U^{(kl)}(\varphi_{kl}(t))=e^{-i\varphi_{kl}(t)
H^{(kl)}}$ 
and 
\be
\varphi_{kl}(t)= \int_0^t g({\bm r}_k(t'),{\bm r}_l(t'))dt'.
\ee
At a  time $t$, the quantum state is fully determined by the $N(N-1)/2$
phases $\varphi_{kl}(t)$, which in turn are determined by the interaction history of the
$N$ particles. Each phase can be interpreted as a matrix element  
$\Gamma_{kl}=\varphi_{kl}$ of  an
adjacency matrix $\Gamma(t)$ defining a weighted graph. Thus,
many entanglement properties of the state can be expressed in 
simple graph--theoretical terms.

We now focus our attention on the case of an Ising-type interaction with 
$H^{(kl)}=\ket{11}_{kl}\!\bra{11}$ \cite{He03}.
We further assume that  all particles are initially prepared in the internal state 
$|+\rangle=1/\sqrt{2}(|0\rangle + |1\rangle)$, and interact only
when they collide, i.e.,  $g(r_{kl})=g_o$ for  $r_{kl}(t)= |{\bm r}_k(t) - {\bm r}_l(t)| \leq d_o$ and $g(r_{kl})=0$ otherwise. These restrictions are made 
for simplicity, but similar methods 
can be applied to general commuting $H^{(kl)}$, arbitrary
$ g({\bm r}_k(t),{\bm r}_l(t))$ and pure separable initial states (without additional overhead).

The evolution of the initial state $|\Psi_0\rangle=\ket{+}^{\otimes N}$ can be
straightforwardly described in the standard basis
$\{\ket{0},\ket{1}\}^{\otimes N}$,
\begin{equation}
U_t \ket{+}^{\otimes N}=2^{-\frac{N}{2}} \sum_{s} U_t
\ket{\mathbf{s}}=2^{-\frac{N}{2}}\sum_s e^{i\frac{1}{2}
\mathbf{s}\cdot\Gamma(t)\cdot\mathbf{s}}\ket{\mathbf{s}}\mbox{,}
\label{eq:Uphase}
\end{equation}
where  the sum is carried out over all $N$-digit binary vectors
$\mathbf{s}$, i.e., over all $2^N$ different combinations of zeros and
ones. We note here that all the time dependence is in the adjacency matrix $\Gamma(t)$ of the graph. The parametrization of the quantum state in terms of a weighted graph that summarizes the ``collisional history'' of the gas is both intuitive and useful for our computations.

Many properties of this  global pure state can be understood in terms of the reduced density matrices of its subsystems. Since the  unitary operations in \eqref{eq:Us} commute with each other,
the evolution of a set $A$ of $N_A$ particles  can be separated into
two contributions. The first  entangles particles within $A$
and is determined by the block  $\Gamma_{A\!A}$ of the adjacency
matrix. The  second contribution couples the subsystem $A$ to the rest $B$
of the system through the off-diagonal block $\Gamma_{A\!B}$. 
The effect of the latter  can be obtained by tracing out the set of particles $B$ from the state
$\ket{\Psi_t}$:
\begin{eqnarray}
\tilde{\rho}_A&=&
\frac{1}{2^{N}} \tr_B \sum_{s,s'}^{2^N-1} e^{i\frac{1}{2}(
\mathbf{s}.\Gamma.\mathbf{s}-\mathbf{s}'\cdot\Gamma\cdot\mathbf{s}')}
\ketbra{\mathbf{s}}{\mathbf{s}'}  \nonumber \\
&=&\frac{1}{2^{N_A}}\sum_{s_A,s_A'} (\frac{1}{2^{N_B}} \sum_{s_B}e^{i
(\mathbf{s}_A-\mathbf{s}'_A)\cdot\Gamma_{AB}\cdot\mathbf{s}_B})\ketbra{\mathbf{s}_A}{\mathbf{s}'_A}\mbox{
}
\label{rhoA}
\end{eqnarray}
The second equality is obtained by writing
$\ket{s}=\ket{s_A}\ket{s_B}$, and the tilde in $\tilde{\rho}_A$ indicates that interactions 
within subsystem $A$ are not taken into account ($\Gamma_{A\!A}$ is set to zero). Clearly, the values of the block  $\Gamma_{B\!B}$ do not affect the properties of either $\rho_A$ or $\tilde{\rho}_A$ .
In the standard basis, each off-diagonal element (``coherence'') of the initial state is
decreased by a factor, 
$\rho_{s_A {s'}\!\!_A}(t)=C_{s_A {s'}\!\!_A} \rho_{s_A
{s'}\!\!_A}(0)$, while diagonal elements remain untouched. The
multiplying factor can be conveniently written as
\begin{equation}
C_{s_A {s'}\!\!_A} =
e^{i
\frac{1}{2}\sum_k(\mathbf{s}_A-\mathbf{s}_A')\cdot\mathbf{\Gamma_{k}}}
\prod_{k=1}^{N_B}
\cos[\textstyle{\frac{1}{2}}(\mathbf{s}_A-\mathbf{s}_A')\cdot\mathbf{\Gamma_{k}}]
\label{eq:HadProd}
\end{equation}
where we have defined the $N_A$-dimensional vector
$(\mathbf{\Gamma_{k}})_j=\Gamma_{k j}$ for each particle $k\in B$. 
In this form, we see that the total effect of the interactions with particles in $B$ on a particular coherence $C_{s_A {s'}\!\!_A}$ of $\rho_A$ can be obtained by multiplying the effects of each individual particle in $B$.  More succinctly, if
$\rho_A^{(k)}$ is the state of the subsystem due to the sole effect
of particle $k\in B$, then  the state $\rho_A$ is obtained (up to
normalization) by the \emph{Hadamard product} of all
$\{\rho_A^{(k)}\}_{k=1}^{N_B}$ written in the standard basis, that is
by their component-wise multiplication. 
This observation can also be understood within the context of  Valence
Bond Solids (VBS) as was recently  shown in \cite{Du04}. 
The decomposition into Hadamard products allows one to
read off the matrix elements of $\rho_A$ from the adjacency matrix, but 
most importantly it signifies that one can efficiently compute
reduced density operators $\rho_{A}$ of small subsystems $A$, even
when the size $N$ of the total system  is essentially arbitrarily
large. The computational effort scales only linearly with $|B|$ in contrast
 to the general case where   the computational resources to calculate $\rho_A$ scale 
  exponentially with $|B|$  (because the partial trace has to be performed over all $2^{|B|}$ basis
states in $B$). 

The  time dependence of the quantum state of the system 
in terms of  $\Gamma(t)$ \eqref{eq:Uphase}, together with the efficient method \eqref{eq:HadProd} to compute the state of (small) sub-systems are  crucial properties that allow us to study 
spin gases. For a complete characterization of the dynamics, one still needs to find, for each particular gas model, the behavior of the stochastic function $\Gamma(t)$.
This task amounts to assigning a probability $p_{\Gamma(t)}$ to every collisional history $\Gamma(t)$. However, in order to  calculate the evolution of average properties, it is enough to have the probability distribution of the $\Gamma$'s at every given time $p_t (\Gamma)$.

{\bf Entanglement and decoherence.} While entanglement of bipartite systems is rather well understood, entanglement properties of multipartite systems are in general difficult to determine.  
 However,  for pure global states, we can get a broad picture of the entanglement in a multipartite system by considering all possible splits of the set of parties in two groups ($A$ and $B$), and analyzing  their bipartite  entanglement.
 For each of the $2^{N-1}$ bipartitions, the entanglement properties can be
completely determined from the Schmidt decomposition, $|\Psi_t\rangle
=\sum_k \sqrt{\lambda_k} |k\rangle_A|k\rangle_B$, where the
$\lambda_k$ are the eigenvalues of the reduced density matrix
$\rho_A = \tr_B|\Psi_t\rangle\langle\Psi_t|$. The entropy of entanglement $S_{\bm
A}=\tr(\rho_A \log_2 \rho_A)$, i.e., the von Neumann
entropy of the reduced density operator $\rho_A$, provides a suitable
measure of bipartite entanglement, quantifying bipartite aspects of
multipartite entanglement. However, as mentioned before, the calculation of
reduced density matrices is, in general, very difficult  if not impossible 
(exponential scaling in both $N_A$ and $N_B$). 
Even the simple criterion to ascertain whether the system is entangled or not
(rank$(\rho_A) >1$ or $=1$) may be impossible to check. 
For states $|\Psi_t\rangle$ which occur in the spin gases under
consideration, many of these restrictions do not apply. First,
we can use \eqref{eq:HadProd} to determine in an
efficient way  the density matrix $\rho_A$ of small subsystems $A$ and hence calculate the entropy of entanglement with respect to all such bipartitions of the system. 
Interactions within $A$ do not change  the entanglement with the rest of the system and can therefore be ignored.  We can also neglect the phase factors in 
   $C_{s s'}$ \eqref{eq:HadProd}, since they can be cancelled
    by applying  local unitaries $V^{(j)}_{kl}=\delta_{kl}\exp(-i \frac{1}{2} k \sum_m \Gamma_{j m})$
   on every qubit $j\in A$.
 Second,  we have a simple criterion for 
the presence of entanglement \footnote{This can be easily proven by, e.g. , checking the equivalent condition that the product in \eqref{eq:HadProd} is less than one for some $C_{s s'}$. }: The state
$|\Psi_t\rangle$ is entangled with respect to the partition $A\!-\!B$ \emph{iff} the two groups are connected (i.e. an interaction between some particle in $A$ and some particle in
$B$ has taken place).  We stress  that since we can  compute  $\rho_A$ of small-sized subsystem, we can also compute quantities such as  the multipartite Meyer-Wallach pure-state entanglement measure  \cite{Me02}, which only depends on single-body density matrices, or  the ``correlation strengths" \cite{Asch04} for finite blocks, which depend on the reduced density matrices of the block and all of its reductions, or classical correlation functions like those used  in generalized $n$-party  Bell-inequalities \cite{werner01}.

A different approach to describe global aspects of the multipartite entanglement  present in the gas is
to ask whether or not entanglement can be created, or \emph{localized}, between two arbitrary subsets of particles $A_1\!-\!A_2$ by performing \emph{local} operations on the other particles. 
Again, we find that for the class of states that emerge in the spin gas 
there is a very simple criterion to answer this question: entanglement between $A_1\!-\!A_2$ can be created by local operations \emph{iff} there exists an path between $A_1$ and $A_2$ in the corresponding graph. The necessity of the condition is obvious, while the sufficiency  follows from these two facts: (i) Collisions where particle $k$ is involved  can be undone (up to local operations) 
 by measuring $k$ in the $z$-basis $\{\ket{0},\ket{1}\}$.
(ii)  If some particle $i$ is connected to a particle $j$ via an intermediate particle $k$ with generic phases 
$\varphi_{ik}$ and $\varphi_{kj}$, then
a projection onto a state $\ket{\tilde{x}}_k \notin \{\ket{0},\ket{1} \}$ effectively
creates an entangling operation on the  particles $i$ and $j$ (unless both are in one of the states of the standard basis). One can hence perform measurements (e.g. along the $x$-axis) on the particles found in the connecting path, $z$-measurements on the rest, and ends up with an entangled pair. Determining whether two particles in a graph are connected is known as the reachability problem, which  can be solved by an algorithm of $O(N^2)$. 
One can try to optimize this procedure to obtain the maximum average entanglement, called \emph{localizable entanglement}.
This quantity, which has recently been introduced  in \cite{Ve03}, is bounded from below by the two-point classical correlation function, and from above by  the entanglement of assistance. Both bounds can be readily calculated from the reduced density matrix $\rho_{A_1A_2}$ of the subsystems and can therefore  be computed efficiently for the spin gas.

\bigskip 

Let us now adopt a slightly different point of view that will prove useful in explaining some central features of our models. We focus our attention on a particular set of (probe) particles, and study the map ${\cal E}_t(\rho)$  induced by their collisions with the other particles, which we call the background gas.  On the one hand, the action of this map on the state $\bigotimes_{j\in A}\ket{+}_j$ gives  $\tilde{\rho}_A$, which provides us with the necessary information to calculate the entanglement between $A$ and the rest of the system. On the other hand, the action of the map can be understood as a decoherence mechanism acting on general states of probe particles. 
 
 A map is fully specified by its action on an operator-basis. In particular, it is straightforward to show 
 that ${\cal E}_t(\ketbra{s}{s'})=C_{s s'}\ketbra{s}{s'}$, with $C_{s s'}$ given by \eqref{eq:HadProd}, as long as the state $\rho_B$ of the background particles fulfills $\tr(\ket{1}_k\!\!\bra{1}\rho_B)=1/2$ for all   $k\in B$ \footnote{The extension to an arbitrary state of the background gas is direct as long as its diagonal matrix elements in the standard basis are known.}. 
This means that in a possible physical realization of such decoherence studies one can relax the experimental conditions that would be necessary to keep the background gas in a coherent state.
In addition, the Hadamard product structure of the maps has the same implications that we saw for states \footnote{ The VBS picture can thus be extended to maps acting on a subset $A$ by  taking the Hadamard product of enlarged states $E_t^{l}= (\eins_{A'}\otimes {\cal E}^{l}_{A}) \ketbrad{\Phi}$ where $\ket{\Phi} = \otimes_{k=1}^{|A|}\ket{\phi^+}_{k' k}$, $\ket{\phi^+} =1/\sqrt{2}(\ket{00}_{A'\!A} + \ket{11}_{A'\!A})$, and ${\cal E}^{l}_{A}$  is the action of particle $l$. 
One can then easily obtain the action of the total map by using the  isomorphism between states and completely positive maps \cite{Ci00}.}.
Semi--quantal gases provide therefore an interesting model for  decoherence in
 a mesoscopic environment. These models can be treated exactly and
 can be highly non-Markovian (i.e., show memory effects) in some regimes.
 
 From  \eqref{eq:HadProd} we find that if a part $A1$ of the probe subsystem $A$ does not have any common collisional partners with the rest  $A2$ of the subsystem  then $C_{s_A,{s'}_{\!\! A}}=C_{s_{A1},{s'}_{\!\! A1}} C_{s_{A2},{s'}_{\!\! A2}}$ where $\mathbf{s}_A=(\mathbf{s}_{A1},\mathbf{s}_{A2})$. Accordingly, we will say that the maps or channels acting on each part of the sub-system are \emph{independent} (or uncorrelated) \footnote{However,  there might actually be small correlations between the channels because the presence of a particle in  the neighborhood of a probe particle might change the probabilities for a second particle to collide in a different part of the subsystem}.
On the other hand,  if two parts of the subsystem share collisional partners
then some coherences will be nearly unaffected by such correlated collisions,  while others will suffer an increased decay (as compared to uncorrelated collisions). 
For instance, consider the case were  two probe particles ($1$ and $2$) 
have a very  similar collision pattern, i.e., $\Gamma_{2j}= \!\Gamma_{1j}\!+\!\delta_j$ for all $j$,
then it follows that coherences associated with $\ketbra{01}{10}$ will only decay by a factor $2^{-N_B} \sum_{s_B}e^{i\mathbf{\delta}\cdot\mathbf{s}_B}$, while $\ketbra{11}{00}$  will be ``super-damped'' by $2^{-N_B} \sum_{s_B}e^{i (2 \mathbf{\Gamma_1}+\mathbf{\delta})\cdot\mathbf{s}_B}$. 

The above conclusions follow directly from the values of the adjacency matrix at a given time.
We now proceed to study the explicit time dependence of the map, where
 we concentrate on one central aspect: Markovian versus non-Markovian dynamics. 
Both effects can already be seen in the evolution of a single probe particle, and essentially correspond to the following two types of collision patterns that occur during a short time interval $\Delta t$:
(i) Markovian:  At every time step $\delta t$ the probe particle collides with a different particle and 
accumulates a small interaction phase $\delta\varphi $. The state of the probe particle will then decohere exponentially fast with the number of time steps $k=\Delta t /\delta t$: $|\rho_{01}|=[\cos(\delta\varphi/2)]^k=e^{-\Delta t/\tau_e }$ with $\tau_e\approx 8   \delta t/\delta\varphi^{2}$.
(ii) Non-Markovian (or coherent coupling):  In the time interval $\Delta t$ a given gas particle has collided $k$ times with the probe. The coherent addition of the interaction phase leads to a Gaussian type of decay:   $|\rho_{01}|=\cos(k \delta\varphi/2)\approx e^{- \Delta t^2/(2\tau_g^2)}$ with $\tau_g=2 \delta t/\delta\varphi$.
Additionally, if we assume no control over quantum or classical degrees of freedom of the background gas, as in our the decoherence analysis, we should average the effect of the maps  ${\cal E}_t(\rho)$ over all possible collision patterns at the given time: 
$\bar{C}_{s,s'}(t)=\int \mathrm{d}\Gamma p_t(\Gamma) C_{s,s'}(\Gamma)$ where $p_t(\Gamma)$ is the probability that at a time $t$ the adjacency matrix is  $\Gamma$.
So, the phase factor in  \eqref{rhoA} suffers two types of averaging effects, one over collision histories, and the other one over all possible combinations of excited (i.e. state $\ket{1}$) collisional partners, i.e. over all $\mathbf{s}_B$. A complete characterization of the dynamics of the probe can be given in terms  of the overall probability distribution $p_{\mathbf{z}_A}(\epsilon)$. Here, $\epsilon=\mathbf{z}_A\!\cdot\!\Gamma_{AB}\!\cdot\!\mathbf{s}_B$ is the phase in \eqref{rhoA} for a given coherence specified by binary vector $\mathbf{z}_A=\mathbf{s}_A-\mathbf{s}'_A$. One then finds that the coherence falls off essentially with the width $\sigma_\Gamma(t)$ of the distribution as $\exp(-\sigma_\Gamma(t)^2/2)$ and acquires a phase $\Phi\sim \langle \epsilon\rangle$. The two extreme regimes mentioned before correspond to $\sigma_\Gamma(t)\propto \sqrt{t}$ (Markovian) and $\sigma_\Gamma(t) \propto t$ (Non-Markovian). We point out, however, that the dynamics will  typically be very rich showing combinations of both effects and  further non-trivial features such as finite--size effects.

{\bf Boltzmann gas.}
 We consider a dilute ideal gas of $N$ particles in thermal equilibrium with a mean free-path comparable to the size of the enclosing volume.  The  statistical state of the gas is fully specified by the density $n$, the volume $V$, and the temperature $T$.  We assume Stosszahlansatz (or molecular chaos) and hence take  a homogeneous and uncorrelated spatial distribution of the particles (density), and an uncorrelated velocity distribution. The latter is  given by the  Maxwell-Boltzmann distribution and is characterized by the single parameter $\sigma=\sqrt{k_B T /m}$, where $m$ is the mass of the particles and $k_B$ the Boltzmann constant.
We further assume a hard-sphere model for collisions between particles of diameter $d$ and that at every collision particles acquire a phase inversely proportional to their relative velocity, $\varphi_{kl}=\gamma / v_{kl}$.  We study the entanglement that arises in the system if at a given time $t=0$ the internal state  of all particles is initialized to  $\ket{+}$.

One could compute entanglement properties by  direct simulation of the Boltzmann gas. Here,  however, we will focus on regimes where analytical results can be obtained, namely for  large collisional phases and arbitrary times, or  for arbitrary phases and in the limits of short and infinite times. 
In what follows we will  use the von Neumann entropy as well as the R\'eyni entropy as a measure of pure state entanglement. R\'enyi entropies, $S_q(\rho)=(1-q)^{-1}\log_2(\tr\rho^q)$, for $q>0$ are known to be non-increasing functions of their parameter $q$, i.e. $S_q\leq S_{q'}$ for $q'>q$. Moreover, in the limit $q\! \rightarrow\!1$ the R\'enyi entropy coincides with the von Neumann entropy, and therefore $S_{q= 2}=-\log_2(\tr\rho^2)$ provides a lower bound to the von Neumann entropy $S(\rho)=S_{q\rightarrow 1} \geq S_{q= 2}$.

In a regime of large collisional phases $\phi\sim \gamma \sigma^{-1}\gg 1$ (i.e., large interaction constant  or low temperatures ) we can assign to each collision event a random phase in $[0,2 \pi]$. This already allows us to find the expected entropy for short times $r t < 1$, where $r= \pi d^2 n \mean{v_r}$ is the collision rate  and  $ \mean{v_r}=\sqrt{16 K T (m\pi)^{-1}}$ is the mean relative velocity. For these short  times a particle will typically collide at most once (with probability $r t$), and  an expected entropy of $\frac{1}{2\pi}\int S(\phi) \mathrm{d}\phi=2-\log_2 e$.  Hence, we find $\mean{S_{1}}=r t (2-\log_2 e) + \mathcal{O}((rt)^2 )$.   The entropy of entanglement  between a block $A$ of size $N_A$ and the rest of the system $B$ can be readily obtained by 
counting the collisions that typically occur between particles of $A$ and $B$: 
\begin{equation}
\mean{S_A}\approx \frac{N_A N_B}{N-1} r t (2-\log_2 e) \mbox{  for }  r t < 1. 
\label{SA}
\end{equation}

For arbitrary times, we can 
use the convexity of the logarithm function to obtain a lower bound for the von Neumann and R\'enyi entropies:  $\mean{S_A}\ge S_{q=2}(\rho_A)=- \mean{\log_2(\tr\rho_A^2)}\ge -\log_2(\mean{\tr\tilde{\rho}_A^2})=\log_2 (\sum_{s_A,{s'}_{\!\! A}} \mean{|C_{s_A,{s'}_{\!\! A}}|^2})$. From \eqref{eq:HadProd} we notice that the coherence  $C_{s_A,{s'}_{\!\! A}}$ only depends on the difference $\mathbf{z}_A=\mathbf{s}_A-{\mathbf{s}'}_{\!\! A}$. And in particular, the average  $\mean{|C_{s_A,{s'}_{\!\! A}}|^2}$  depends only on the number $Z_A$ of non-zero entries of $\mathbf{z}_A$.  Each particle $k$ in $B$ will contribute with a factor $1/2$ to the product in \eqref{eq:HadProd}  if it has collided with at least one particle of the subset of $A$ where $\mathbf{z}_A$ has  non--zero entries, while a factor one appears otherwise. Since the probability that no such collision occurs is  $p_{z_A}=\exp(-r t Z_A/N)$,  on average each term in the product will contribute with a factor $1/2 (1-p_{z_A})+ p_{z_A}=(1+\exp(-r t Z_A/N))/2$, and taking into account combinatoric factors we arrive at
\begin{equation}
\mean{S_A(t)}\geq -\log_2\left(\frac{1}{2^{N}}\sum_{z_A=0}^{N_A}{N_A \choose Z_A}(1+e^{-rt\frac{Z_A}{N-1}})^{N_B}\right)\mbox{.}
\label{eq:genSA}
\end{equation}
Numerical results for  $N_A\leq 8$ and arbitrary system sizes show that this lower bound is also a good estimate and describes well the behavior of the entropic entanglement.  This lower bound can be supplemented by the upper bound $\mean{S_A}\leq N_A \mean{S_1}$. The latter follows from  the sub-additivity property of the von Neumann entropy.  

 In the short- or long--time limits we can simplify the above expression for the entanglement between two arbitrary parts of the Boltzmann gas. For short times $r t N_A/(N-1)<1$, we find
 \begin{equation}
 \mean{S_A(t)}\geq - \log_2(1-\frac{N_A N_B}{4 (N-1)} rt)\approx \frac{N_A N_B}{4 \ln 2 (N-1)} rt\mbox{,}\nonumber
 \end{equation}
 which is consistent with the exact result \eqref{SA} for short times.
 
In the long--time limit all particles will have collided with all other particles many times and accumulated phases $\varphi_{kl}\gg 1$, independently of the collisional phase per collision. We refer to such a state as the \emph{equilibrium} state.
As times  increases, $ r t  \gg N $,  every term with $z_A>0$ in the sum  \eqref{eq:genSA} approaches one (disregarding the binomial factor) while the $z_A=0$ term is equal to two, 
\begin{equation}
\mean{S_A}\geq -\log_2\left(\frac{1}{2^{N_A}}+\frac{1}{2^{N_B}}-\frac{1}{2^{N}} + \textstyle{\frac{N_A N_B }{2^{N}}}e^\frac{-r t}{N-1}\right)
\nonumber
\end{equation}
The equilibrium state ($rt\rightarrow \infty$) has thus the interesting feature that $|\Psi_\infty\rangle$ is maximally entangled with respect to {\em all} possible bipartitions, i.e. $S_A\approx N_A$, provided that the total number of particles $N$ in the gas is sufficiently large.  This is a non-trivial statement especially in the case where both partitions are similarly large. We find that, for whatever bipartition one takes, the expected entropy of entanglement is at most a single bit away from its maximal value: $N_A\geq\mean{S_A}>N_A-1$.
  This result is in agreement with the findings of Page \cite{page93} and subsequent work studying average or typical entanglement properties of the whole set of  multipartite pure states. 
  Another remarkable  property of the equilibrium state is that the localizable entanglement between \emph{any} pair of atoms approaches its maximum value of one e-bit as $N$ increases. The proof is simple. For any pair of atoms $\{i,j\}$ with an interaction phase $\varphi_{ij}$  we can find a third particle $k$  with which both atoms have accumulated a phase $\varphi_{ik}\approx \varphi_{jk}\approx \pi$. Then, we measure the remaining  gas particles in the $z$-basis (to  effectively  decouple them)  and  particle $k$  in the $x$-basis.
 A straightforward calculation shows that  the resulting state of particles $\{i,j\}$ is always maximally entangled  \footnote{The distance from the resulting state to a maximally entangled state is of the order of the deviations  of phases $\delta\varphi_{ik}= \pi - \varphi_{ik}\sim  2\pi/N$. 
For finite $N$, the proposed strategy is not optimal.}, no matter  what the initial $\varphi_{ij}$ was.
 Whereas the dynamics of the entanglement strongly depends on the particular type  and regime of the semi--quantal gas, the equilibrium state will,  in general,  be of the type we just described. In particular, it describes the state of the semi-quantal Boltzmann gas also in the regime of  small collisional phases, and the lattice gas studied below.

To conclude our analysis of the semi-quantal Boltzmann gas,  let us give the short--time entanglement in the regime of small collisional phases.  In this regime, we have to take into account the relative velocity of the colliding partners. A simple kinetic theory calculation shows that at short times ($ r t\ll1$)  the expected value for the  squared modulus of the coherence is given by,
 \begin{equation} 
\mean{|C_{01}|^2}_t\approx 1-t n  \pi d^2 \int\!\mathrm{d}v \, p_r(v) v\,  (1-|C_{v}|^2)=1- \alpha t \mbox{,} \nonumber
\end{equation} 
where $p_r(v)$ is the relative velocity distribution and $|C_v|^2=\cos^2(\phi(v)/2)$ is the squared modulus of the  velocity dependent coherence. For small collisional phases $\gamma\sigma^{-1}<1$ the proportionality  factor $\alpha$ can be approximated by
 \begin{equation}
\alpha=4\pi^2 d^2 n (\frac{1}{4 \pi \sigma^2})^{\frac{3}{2}} \int\!\mathrm{d}v \,v^3 \mathrm{e}^{-\frac{v^2}{4 \sigma^2}}\sin^2(\frac{\gamma}{2 v})\approx \frac{1}{4} n \sqrt{\pi} d^2 \frac{\gamma^2}{\sigma} \mbox{.}
\nonumber
\end{equation} 
Following the previous reasoning we find 
\begin{equation}
\mean{S_A(t)}\geq -\frac{N_A N_B}{N-1} \log_2(1-\frac{\alpha}{2} t)\approx \frac{\alpha} {2\ln 2} \frac{N_A N_B}{N-1}  t \mbox{.}
\end{equation}
Thus, by writing $\alpha\approx \frac{1}{4}\sqrt{m \pi} d^2 \gamma^2 \frac{n}{\sqrt{k_B T}}$ we obtain the  rate of entanglement generation  in terms of the thermodynamical variables. Whereas in the low temperature regime entanglement grows with the rate of collisions ($\propto \sqrt{T}$), here we find that entanglement generation is governed by the slow collision events (larger phases) leading to the opposite behavior ($\alpha \propto \sqrt{T^{-1}}$).

{\bf Lattice gas.}  The quantum properties of the system are directly linked to the classical statistical properties of the 
gas through $\Gamma(t)$.Thus,  in general, it is necessary to know the classical $n$-body distributions to give a complete description of the quantum state. For some gas models and regimes (like the Boltzmann gas presented above) correlations play a minor role and one can find analytical descriptions for single-particle phase--space distributions.   In the remainder of this paper  we study  a lattice gas model that exhibits strong correlations. We use numerical simulations  to recreate such classical correlations --- or, in other words, produce  different random realizations of $\Gamma(t)$.
The model has the additional feature that it  can be possibly implemented in a quantum optical system. It has already been shown that an optical lattice can be used to store ultra-cold atomic gases. The degree of control in these experiments is extraordinary.  Various  system parameters can be adjusted ---from the structure of the underlying periodic lattice to the interactions between the atoms---,  opening the door to a wide range of experiments and theoretical proposals \cite{greiner02,mandel03,paredes04, jaksch98, hofstetter02, duan03, damski03}. In particular, one can choose a parameter regime where each lattice site is occupied by at most one atom \cite{jaksch98, greiner02}.  The internal quantum state of  the atom (e.g. two meta-stable hyperfine states) can be stored in coherent superpositions over long time-scales (few minutes), while coherent inter-atomic interactions have also been achieved by cold collisions \cite{mandel03}. These interactions correspond precisely to the  Ising--type  chosen here. One can also find schemes \cite{expref} to induce a random (incoherent) hopping of the atoms from one site to its neighboring sites. Hence, we consider an $M\!\times\!M$ lattice containing $N$ particles that  randomly hop from site to site with a hopping rate  $\eta$, and have nearest--neighbor interactions  with  coupling constant $g_o$ .
 
 We use the above lattice gas model  to investigate the typical multipartite entangled states that arise.  
 An interesting observed effect is the creation of clusters,   that is,  sets of particles that are connected in the graph,  such that  entanglement can be localized between any of the constituent particles.
 One finds that the average number $N_C$ of particles that form
a cluster grows exponentially with time. Similarly, the expected value for
 the maximal distance $\ell_{\rm max}$ between (localizable) entangled 
  particles increases. After a finite time $t_0$, which only depends on the filling factor $\nu=N M^{-2}$  and hopping rate $\eta$, almost all particles are connected. That is,  after this time $t_0$, all particles are
pairwise (localizable) entangled  and hence $\ell_{\rm max} \sim m $,
suggesting  a (bond) percolation phenomenon in the underlying graph. 

\begin{figure}[ht]
\begin{picture}(230,85)
\put(-5,0){\epsfxsize=110pt\epsffile[30 39 706 530]{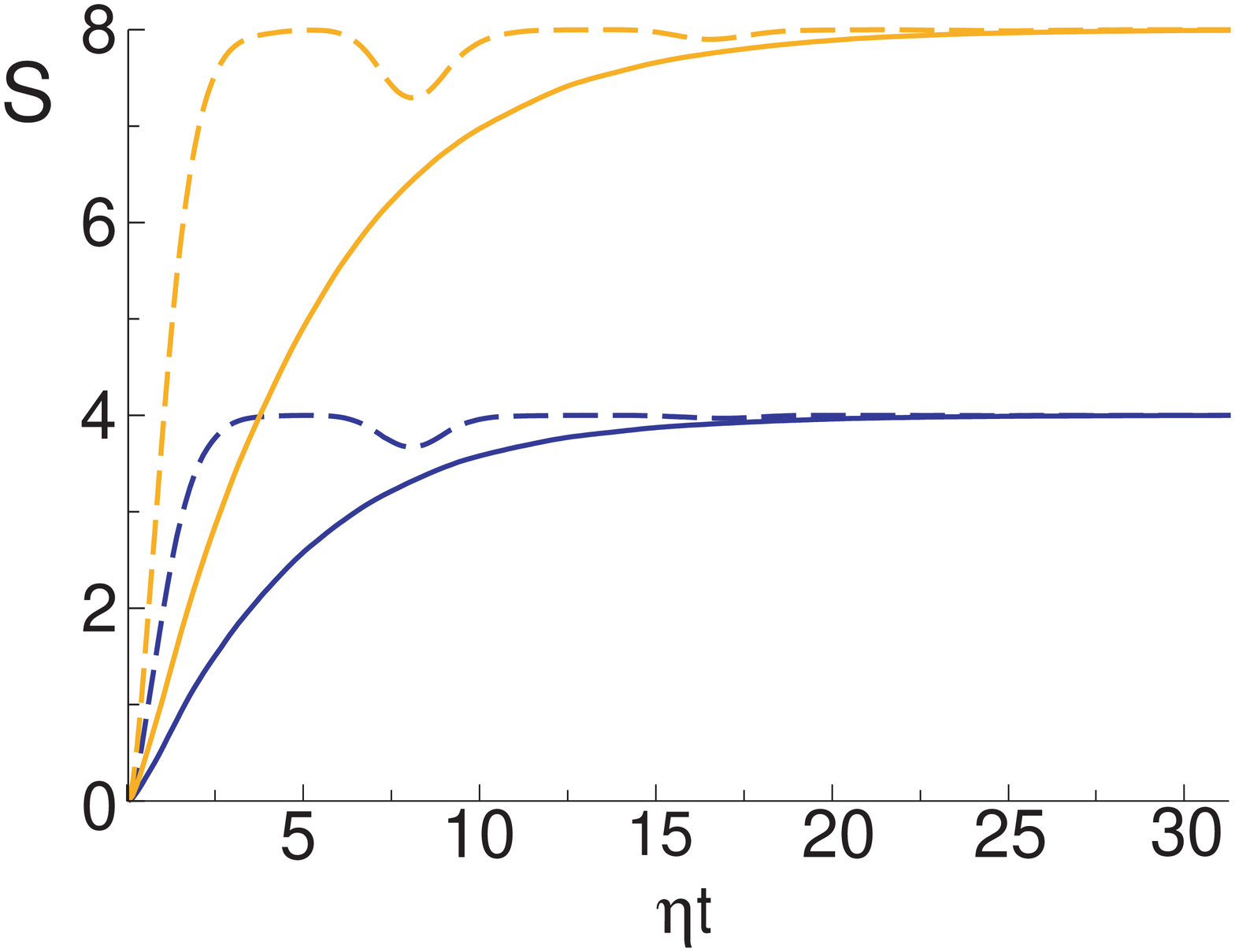}}
\put(115,0){\epsfxsize=110pt\epsffile[30 39 706 530]{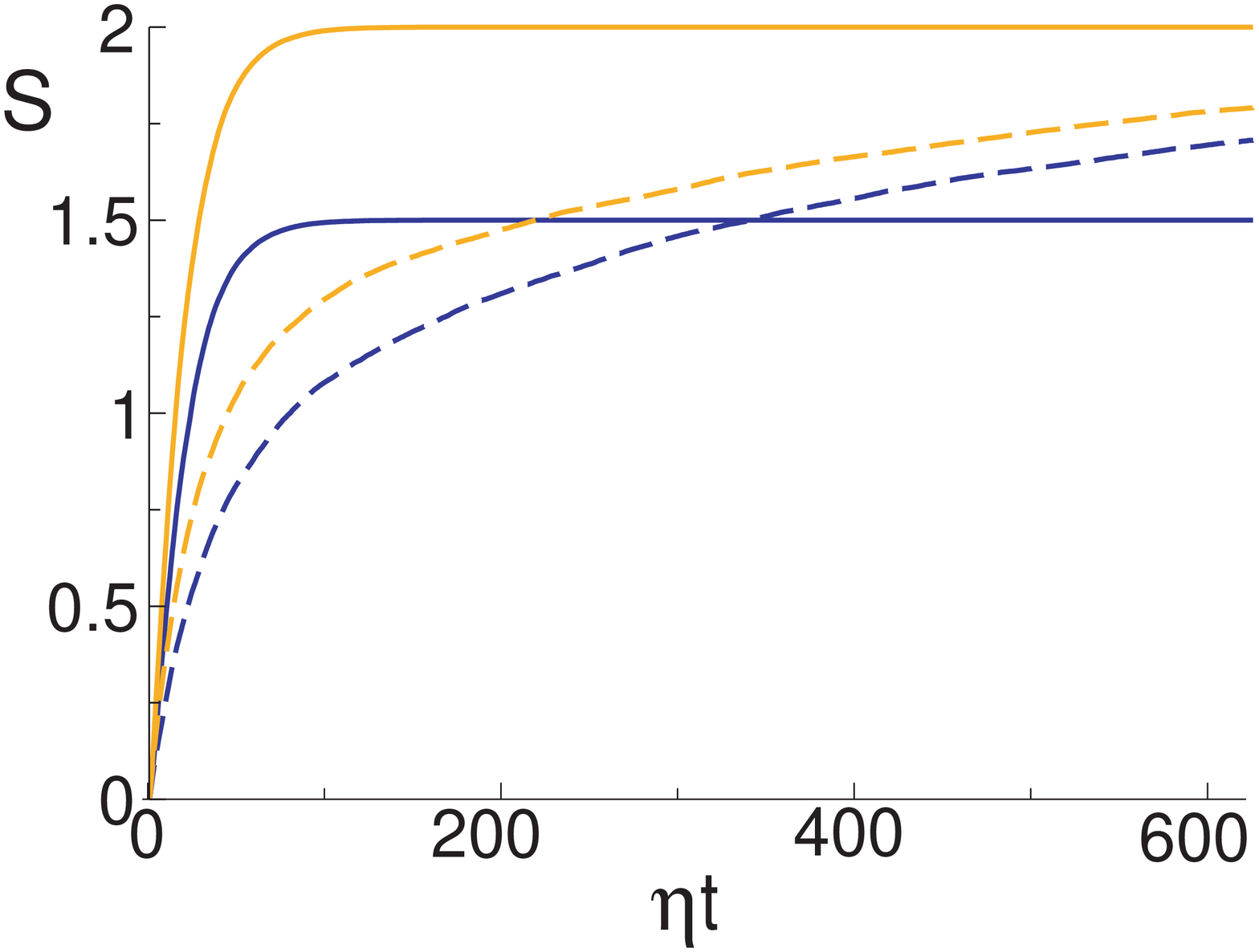}}
\end{picture}
\caption[]{\label{Entanglement} Lattice gas with $M=20$, $g_o= 0.8 \eta$. (a)
Average entropy of entanglement for block sizes $|A|=4$ (dark) and  $|A|=8$ (light)
   and filling factors $\nu=0.9 $ (dashed) and $\nu=0.25$ (solid). (b) Average entropy of entanglement
   for two probe particles that are initially located at a relative distance $\ell=0$ (thin-dark) and $\ell=8$ (thick-light)    and have hopping rates $\eta_p=0$ (dashed)  and $\eta_p=\frac{1}{5} \eta$ (solid).}
\end{figure}
 
Concerning block-wise entanglement, Figure \ref{Entanglement}(a) shows the expected entanglement between a set of randomly chosen particles and the rest of the system as a function of time.
 Since all the density matrix elements are initially equal, $|\Psi_0\rangle=\ket{+}^{\otimes N}$, and  all coherences decay to zero with time, we find that the entanglement saturates to its maximum value. At high filling factors the entropy of entanglement undergoes periodic decreases every time the typical accumulated phase $\phi_{1k}(t)$ with a nearest neighbor reaches $2\pi$, i.e., at $t \approx k 2\pi/ g$ with integer $k$. At lower filling factors we can observe that the entanglement is not additive when increasing the block-size, which  indicates the presence of correlated collisions.  To highlight  this effect, we consider the case where two probe particles move freely in the lattice ---with a hopping rate $\eta_p$ possibly different from the hopping rate 
of the background particles---  and only interact with the gas particles that are in the same position \footnote{A possible implementation could consist of  three different layers of a 3-D optical lattice: the middle layer for the gas, sandwiched between the layers where each of the probes moves independently. We can choose probes  of a different species to selectively change their hopping rate.}.
Fig.\ref{Entanglement}(b) shows the entanglement of the probes with the rest of the system for different
initial relative distances $\ell $ between the two probes, and for different probe hopping rates $\eta_p$. Small probe hopping rates favor repeated collisions with the same particle ---  as opposed  to independent collisions with different particles. As we saw before, this amounts to coherent addition of the interaction phase, which  leads to a much higher rate of entanglement generation.
At $\ell=0$ and fixed probes ($\eta_p=0$), all the collisions with gas particles are exactly correlated. From the above discussion we know that under these circumstances 
the coherences of the type $\rho_{01,10}$ remain untouched while all others decay to zero,
leading to a maximal entropy of $S(\infty)=1.5$.
When the probe particles are given a non-zero hopping rate, they start suffering independent collision events which eventually turn the reduced density matrix into a completely mixed one ($S=2$).

\begin{figure}[ht]
\begin{picture}(230,80)
\put(-10,0){\epsfxsize=125pt\epsffile[0 0 497 318]{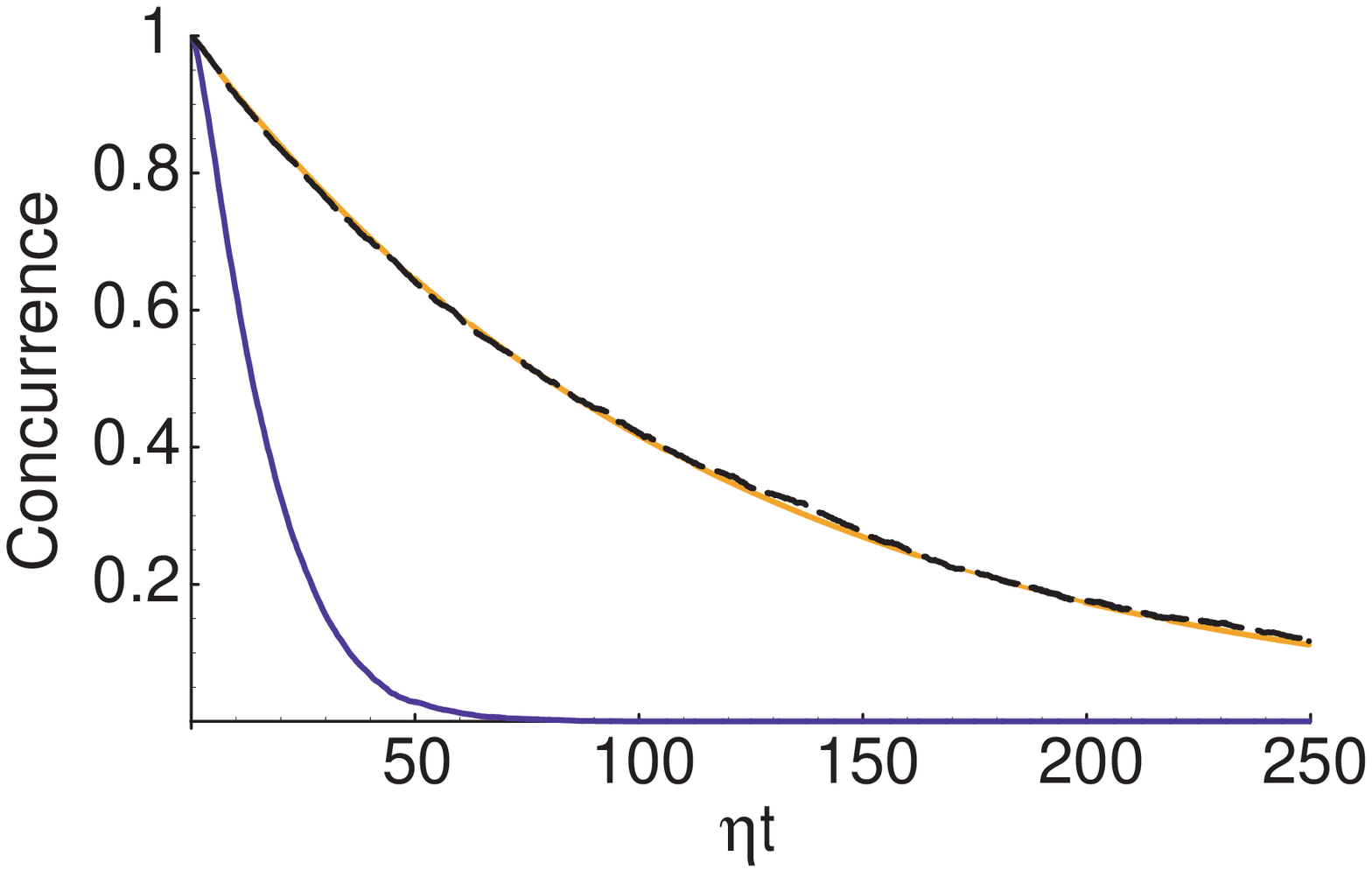}}
\put(90,0){\epsfxsize=125pt\epsffile[ 0 0 497 318]{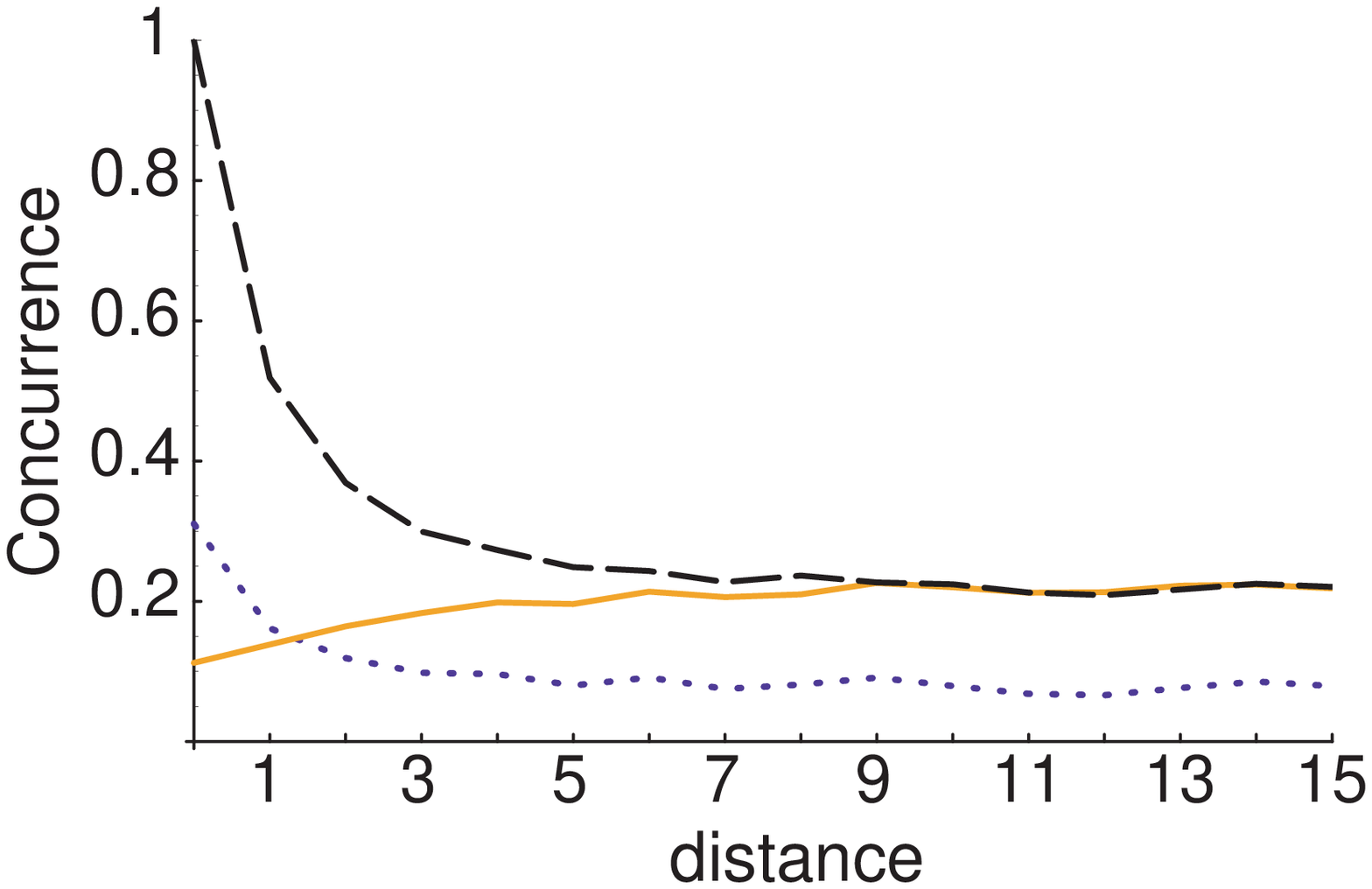}}
\end{picture}
\caption[]{\label{fig:decoh} (a) Concurrence of two probe particles prepared in a maximally entangled Bell state $\ket{\phi^+}=1/\sqrt{2}(\ket{00}+\ket{11})$  in a  $100\!\times\!3200$ lattice gas  with $N=8\times10^4$  for  $g_o=0.8 \eta $ and fixed probes (solid);  and fast moving probes (dashed). Thin-solid curve corresponds to analytical result obtained for the Markovian regime.  (b) Concurrence of two probe particles at  time $t_o=25 \eta^{-1}$ for different initial entangled states as a function of relative distance: $\ket{\psi^+}=1/\sqrt{2}(\ket{01}+\ket{10})$ (dashed-dark) , $\ket{\phi^+}$ (solid-light), and $\ket{G}=1/\sqrt{2}(\ket{+0}+\ket{-1})$ (dotted). }
\end{figure}

In the previous plots  we assumed that the probe particles were initially prepared in the same quantum state as the background particles.
In the following, we study the decoherence of different entangled initial states of the probe particles due to the action of the background gas.
In other words, we analyze properties of the average state of the probes, 
as opposed to average properties of the resulting states.
 We consider  the situation where the probe particles are displaced at a constant speed $v$ relative to the gas.  We vary the speed $v$ of  the probes as well as the distance $d$ between the probe
particles. By doing so, we can distinguish  two different effects, analogous to those studied in the context of entanglement generation: (i) By decreasing the probe speed, we analyze the effect of multiple interactions with the same particle in contrast to interactions with different (independent) particles.
(ii)  By increasing the distance, we turn from correlated to independent
interactions of the probes with particles in the remaining
system.  
 Fig. \ref{fig:decoh}(a) shows the decay of the entanglement in the Bell state $\ket{\phi^+}$
as measured by the concurrence \cite{wootters98}  in two extreme scenarios: (i) The probe particles are fixed in their initial positions ($v=0$). (ii) The probe speed is chosen large enough ($v/a\gg \eta$, where $a$ is the inter-site spacing) so that a fixed value $\varphi=0.1$ can be assigned  to the collisional phase every time a probe particle crosses an occupied site. These two scenarios  illustrate the difference between Markovian and non-Markovian environments. A large probe speed enforces a perfect Markovian behavior which matches the analytical curve (see \footnote{At every time step, each probe particle interacts with a new background gas particle
with probability $\nu$. Hence, after a number $k$ of time steps, the relevant coherence  is given by $|C_{00,11}|=|C_{0,1}|^2=| \nu \exp(i \delta\phi/2) \cos(\delta\phi/2)+(1-\nu)|^{2 k}$.}).
Fig. \ref{fig:decoh}(b) shows the concurrence at a given time $t_o$ as a function of 
the distance for three different entangled states: two Bell states and a cluster state (see figure caption). For Bell states  the concurrence is equal to the absolute value of
their only non-zero off-diagonal element in the density matrix,
 and therefore Fig. \ref{fig:decoh} provides direct information about the individual coherences.
The figure clearly shows the influence of correlated collisions: coherences $\rho_{01,10}$ (appearing in $\ket{\psi^+}$) are robust against correlated noise, 
 coherences $\rho_{00,11}$ (appearing in $\ket{\phi^+}$) are especially fragile under correlated noise,
 and the remaining coherences decay in the same way under correlated or uncorrelated noise (hence, the weak distance dependence of $\ket{G}$).
From Fig. \ref{fig:decoh}(b)  it is also evident that the environments become 
more and more independent as $d$ increases. 
In the limit of large distances (i.e. independent channels) the 
concurrence of the Bell states and the cluster state are related by 
$C_G=\max\{0,\frac{1}{2}(-1+2\sqrt{C_{\mbox{Bell}}}+C_{\mbox{Bell}})\}$.

In this paper we studied the stochastic generation of entanglement in a spin gas where the classical kinematics of the particles drives the quantum state of the many-body system. This provides a novel scenario to study relationships between classical thermodynamical variables and properties of the quantum state of the system such as  quantum correlations, rate of entanglement generation, saturation values and clustering effects. We have shown how to formalize such systems  under quite general conditions and  examined in some detail the cases of the  Boltzmann and the lattice gas. We have fully characterized the bipartite aspects of the entangled $N$-body state at long times, showing that maximum entanglement is asymptotically reached for all possible bipartitions. The formalism has also enabled us to treat  exactly a microscopic  model for non--Markovian decoherence, or in other words, the time dependent maps describing the decoherence process of a set  of probe particles  that are in contact with the gas.  We have seen that by changing the system parameters one can reach qualitatively different behaviors. The proposed models  open the door to study entanglement dynamics in mesoscopic disordered systems. The recent experimental breakthroughs on cold atoms in  optical lattices \cite{greiner02, mandel03, paredes04} provide promising perspectives for the experimental study of spin gases.
 
We thank  J. Eisert, S.J. van Enk,  O. G\"uhne, M. Lewenstein,  A. Miyake,  M. Plenio, S. Popescu, P. Zoller    and K. Zyckowski for discussions. 
This work was
supported by the \"Osterreichische Akademie der Wissenschaften
through project APART (W.D.), the European Union
(IST-2001-38877,-39227) and the DFG.


\begin{thebibliography}{99}

\bibitem{vidal03}
G. Vidal, Phys. Rev. Lett. {\bf 91}, 147902 (2003);
G. Vidal, Phys. Rev. Lett. {\bf 93}, 040502 (2004).

 
\bibitem{arnesen01}
M. A. Nielsen, Ph.D. thesis, University of New Mexico,
1998.;  M. C. Arnesen, S. Bose, and V. Vedral, Phys. Rev. Lett. {\bf 87} , 017901 (2001).

\bibitem{vidal03b}
G. Vidal, J. I. Latorre, E. Rico, and A. Kitaev, Phys. Rev. Lett. {\bf 90}, 227902 (2003).

\bibitem{osborne02}
T. J. Osborne, M. A. Nielsen,
Phys. Rev. A {\bf 66}, 032110 (2002).

\bibitem{osterloh02}
A. Osterloh, L. Amico, G. Falci, and R. Fazio, Nature {\bf 416}, 608 (2002).

\bibitem{damski03}
B. Damski, J. Zakrzewski, L. Santos, P. Zoller, and M. Lewenstein, 
Phys. Rev. Lett.  {\bf 91}, 080403 (2003). 

\bibitem{Ve03}
F. Verstraete, M. Popp and J. I. Cirac, Phys. Rev. Lett. {\bf 92}, 027901 (2004).

\bibitem{schollwock04}
U. Schollwock, \emph{The density-matrix renormalization group}, 
to appear in Rev. Mod. Phys. (2004).


\bibitem{verstraete04}
 F. Verstraete, J. I. Cirac, cond-mat/0407066.

\bibitem{ziman04}
M. Ziman, P. Stelmachovic, V.  Buzek,
quant-ph/0410161, to appear in Open Systems and Information Dynamics; see also review article
W. Zureck, Rev. Mod. Phys. {\bf 75}, 715 (2003).


\bibitem{He03}
M. Hein, J. Eisert,  H.J. Briegel, Phys. Rev. A {\bf 69}, 062311 (2004).

\bibitem{Du04}
W. D\"ur, L. Hartmann, M. Hein, M. Lewenstein and H.-J. Briegel,
quant-ph/0407075.

\bibitem{Me02}
D.A. Meyer and N.R. Wallach, J. of Math. Phys. {\bf 43}, 4273 (2002);
G.K. Brennen, Quant. Inf. and Comp. {\bf 3}, 619 (2003).

\bibitem{Asch04}
H. Aschauer, J. Calsamiglia, M. Hein, H. J. Briegel,
Quant. Inf. and Comp. {\bf 4}, 383 (2004).

\bibitem{werner01}
M.~Zukowski, \v{C}. Brukner,
Phys.~Rev.~Lett. {\bf 88}, 210401  (2002); R. F. Werner, M. Wolf, 
Phys.~Rev.~A. {\bf 64}, 032112 (2001).

\bibitem{Ci00}
J. I. Cirac, W. D\"ur, B. Kraus and M. Lewenstein, Phys. Rev. Lett. {\bf 86}, 544 (2001).


\bibitem{page93}
D. N. Page, Phys. Rev. Lett. {\bf 71}, 1291 (1993);
P. Hayden, D. W. Leung, A. Winter, quant-ph/0407049.

\bibitem{greiner02}
M. Greiner, O. Mandel, T. Esslinger, T. W. H\"ansch, I. Bloch,  
Nature {\bf 415} , 39 (2002).

\bibitem{mandel03}
O. Mandel, M. Greiner, A. Widera, T. Rom, T. W. H\"ansch, and I. Bloch,
Nature, {\bf 425}, 937-940 (2003).

\bibitem{paredes04}
B. Paredes, A. Widera, V. Murg, O. Mandel, S. Fšlling, I. Cirac,  G. V. Shlyapnikov, T. W. H\"ansch and I. Bloch 
Nature {\bf 429},  277-281 (2004).

\bibitem{jaksch98}
D. Jaksch, C. Bruder, J. I. Cirac, C. W. Gardiner and P. Zoller, 
Phys. Rev. Lett.  {\bf 81}, 3108-311 (1998).

\bibitem{hofstetter02}
W. Hofstetter, J. I. Cirac, P. Zoller, E. Demler, M. D. Lukin, Phys. Rev. Lett. {\bf 89}, 220407 (2002). 

\bibitem{duan03}
L.-M. Duan, E. Demler, and M. D. Lukin, Phys. Rev. Lett. {\bf 91}, 090402 (2003).


\bibitem{expref}
This is joint work with P. Zoller and M. Lewenstein , unpublished.

\bibitem{wootters98}
W.K. Wootters, Phys. Rev. Lett. {\bf 80}, 2245 (1998).









\end{thebibliography}
\end{document}